\input harvmac
\input epsf
\def\Title#1#2{\rightline{#1}\ifx\answ\bigans\nopagenumbers\pageno0\vskip1in
\else\pageno1\vskip.8in\fi \centerline{\titlefont #2}\vskip .5in}

%
%
\ifx\epsfbox\UnDeFiNeD\message{(NO epsf.tex, FIGURES WILL BE IGNORED)}
\def\figin#1{\vskip2in}
\else\message{(FIGURES WILL BE INCLUDED)}\def\figin#1{#1}
\fi
\def\Fig#1{Fig.~\the\figno\xdef#1{Fig.~\the\figno}\global\advance\figno
 by1}
%
%
%
%
\def\ifig#1#2#3#4{
\goodbreak\midinsert
\figin{\centerline{\epsfysize=#4truein\epsfbox{#3}}}
\narrower\narrower\noindent{\footnotefont
{\bf #1:}  #2\par}
\endinsert
}

%
%
\font\ticp=cmcsc10
\def\sq{{\vbox {\hrule height 0.6pt\hbox{\vrule width 0.6pt\hskip 3pt
   \vbox{\vskip 6pt}\hskip 3pt \vrule width 0.6pt}\hrule height 0.6pt}}}

\def\ajou#1&#2(#3){\ \sl#1\bf#2\rm(19#3)}
\def\jou#1&#2(#3){,\ \sl#1\bf#2\rm(19#3)}
\def\hf{{1\over 2}}

\def\frac#1#2{{#1\over#2}}
\def\Rads{{R}}
\def\cald{{\cal D}}

\def\eg{{\it e.g.}}
\def\sqrtG{\sqrt{-G}}
\def\calL{{\cal L}}

\def\calR{{\cal R}}
\def\calo{{\cal O}}
\def\hbmn{{\bar h}_{\mu\nu}}
\def\etamn{{\eta_{\mu\nu}}}
\def\calS{{\cal S}}
\def\barh{{\bar h}}
\def\Sct{{S_{\rm grav}}}
%
\lref\deAli{S.P. de Alwis, ``Brane world scenarios and the cosmological
constant,'' hep-th/0002174.}
\lref\deAlii{S.P. de Alwis, ``Brane worlds, the cosmological constant and
string theory,'' hep-th/0004125.}
\lref\APR{N. Arkani-Hamed, M. Porrati, and L. Randall, in progress.}
\lref\DFGK{O. DeWolfe, D.Z. Freedman, S.S. Gubser, A. Karch, ``Modeling
the fifth-dimension with scalars and gravity,''
hep-th/9909134.}
\lref\FLLN{S.~Forste, Z.~Lalak, S.~Lavignac and H.~P.~Nilles,
``A comment on self-tuning and vanishing cosmological constant in the
brane world,'' hep-th/0002164, {\sl Phys.\ Lett.}  {\bf B481} (2000) 360\semi
``The cosmological constant problem from a brane-world
perspective,'' hep-th/0006139, {\sl JHEP} {\bf 09} (2000) 034.}
\lref\BaKrhol{V. Balasubramanian and P. Kraus, ``Spacetime and the holographic
renormalization group,'' hep-th/9903190, {\sl Phys. Rev. Lett.} {\bf 83}
(1999) 3605.}
\lref\EHMmob{R. Emparan, G.T. Horowitz, R.C. Myers, ``Black holes radiate
mainly on the brane,'' hep-th/0003118, {\sl Phys. Rev. Lett.} {\bf 85} 2000 499.}
\lref\KKOP{P. Kanti, I.I. Kogan, K.A. Olive, and M. Pospelov,
``Single-brane cosmological solutions with a stable compact extra
dimension,'' hep-ph/9912266, {\sl Phys. Rev} {\bf D61} (2000) 106004.}
\lref\EVer{E.~Verlinde,
``On RG-flow and the cosmological constant,''
{\sl Class.\ Quant.\ Grav.}  {\bf 17}, 1277 (2000), hep-th/9912058.}
\lref\VerVer{E.~Verlinde and H.~Verlinde,
``RG-flow, gravity and the cosmological constant,''
{\sl JHEP} {\bf 0005}, 034 (2000), hep-th/9912018.}
\lref\CEGH{C.~Csaki, J.~Erlich, C.~Grojean and T.~Hollowood,
``General properties of the self-tuning domain 
wall approach to the  cosmological constant problem,''
{\sl Nucl.\ Phys.} {\bf B584}, 359 (2000), hep-th/0004133.}
\lref\HLZ{G.~T.~Horowitz, I.~Low and A.~Zee,
``Self-tuning in an outgoing brane wave model,''
hep-th/0004206.}
\lref\KSSii{S.~Kachru, M.~Schulz and E.~Silverstein, 
``Bounds on curved domain walls in 5d gravity,'' 
{\sl Phys.\ Rev.}  {\bf D62}, 085003 (2000), hep-th/0002121.}
\lref\KSS{S.~Kachru, M.~Schulz and E.~Silverstein,
``Self-tuning flat domain walls in 5d gravity and string theory,''
{\sl Phys.\ Rev.}  {\bf D62}, 045021 (2000), hep-th/0001206.}
\lref\ADKS{N.~Arkani-Hamed, S.~Dimopoulos, N.~Kaloper and R.~Sundrum,
``A small cosmological constant from a large extra dimension,''
{\sl Phys.\ Lett.}  {\bf B480}, 193 (2000), hep-th/0001197.}
\lref\Kal{N.~Kaloper, ``Bent domain walls as braneworlds,''
{\sl Phys.\ Rev.}  {\bf D60}, 123506 (1999), hep-th/9905210.}
\lref\CGKT{C.~Csaki, M.~Graesser, C.~Kolda and J.~Terning,
``Cosmology of one extra dimension with localized gravity,''
{\sl Phys.\ Lett.}  {\bf B462}, 34 (1999), hep-ph/9906513.}
\lref\CGRT{C.~Csaki, M.~Graesser, L.~Randall and J.~Terning,
``Cosmology of brane models with radion stabilization,''
{\sl Phys.\ Rev.}  {\bf D62}, 045015 (2000), hep-ph/9911406.}
\lref\ADD{I.~Antoniadis, N.~Arkani-Hamed, S.~Dimopoulos and G.~Dvali,
``New dimensions at a millimeter to a Fermi and superstrings at a TeV,''
{\sl Phys. Lett.} {\bf B436}, 257 (1998) hep-ph/9804398.}
\lref\GKR{S.~B.~Giddings, E.~Katz and L.~Randall,
``Linearized gravity in brane backgrounds,'' hep-th/0002091,
{\sl JHEP} {\bf 0003}, 023 (2000).}
\lref\RSI{L.~Randall and R.~Sundrum,
``A large mass hierarchy from a small extra dimension,'' hep-ph/9905221,
{\sl Phys.\ Rev.\ Lett.}  {\bf 83}, 3370 (1999).}
\lref\RSII{L. Randall and R. Sundrum, ``An alternative to
compactification,'' hep-th/9906064\jou Phys. Rev. Lett. &83 (99) 4690.}
\lref\WittITP{E. Witten, remarks at ITP Santa Barbara Conference ``New
dimensions in field theory and string theory,''
http://www.itp.ucsb.edu/online/susy\_c99/discussion/.}
\lref\Gubs{S.S. Gubser, ``AdS/CFT and gravity,'' hep-th/9912001.}
\lref\Hver{H. Verlinde, ``Holography and compactification,''
hep-th/9906182.}
\lref\LyRa{J. Lykken and L. Randall, ``The shape of gravity,''
hep-th/9908076, {\sl JHEP} {\bf 0006}, 014 (2000).}
\lref\EHM{R. Emparan, G.T. Horowitz, and R.C. Myers, ``Exact description of
black holes on branes,'' hep-th/9911043.}
\lref\EHMii{R.~Emparan, G.~T.~Horowitz and R.~C.~Myers,
``Exact description of black holes on branes. II: Comparison with BTZ
black holes and black strings,'' hep-th/9912135, {\sl JHEP} 
{\bf 0001}, 021 (2000).}
\lref\GaTa{J. Garriga and T. Tanaka, ``Gravity in the brane world,'' 
hep-th/9911055, {\sl Phys.\ Rev.\ Lett.}  {\bf 84}, 2778 (2000).} 
\lref\GoWi{W.D. Goldberger and M.B. Wise, ``Modulus stabilization with bulk
fields,'' hep-ph/9907447\jou Phys. Rev. Lett. &83 (99) 4922.}
\lref\SuWi{L.~Susskind and E.~Witten, 
``The holographic bound in anti-de Sitter space,'' hep-th/9805114.}
\lref\HeSk{M. Henningson and K. Skenderis, ``The holographic Weyl
anomaly,'' hep-th/9806087, {\sl JHEP} {\bf 9807:023} (1998).}
\lref\BaKr{V. Balasubramanian and P. Kraus, ``A stress tensor for anti-de
Sitter gravity'' hep-th/9902121\jou  Commun. Math. Phys. &208 (99) 413.}
\lref\MaNu{J.~Maldacena and C.~Nunez,
``Towards the large N limit of pure N = 1 super Yang Mills,'' hep-th/0008001.}
\lref\HMS{S.~Hawking, J.~Maldacena and A.~Strominger,
``DeSitter entropy, quantum entanglement and AdS/CFT,''
hep-th/0002145.}
\lref\KlSt{I.~R.~Klebanov and M.~J.~Strassler,
``Supergravity and a confining gauge theory: Duality cascades and  
$\chi$SB-resolution of naked singularities,''
hep-th/0007191.}
\lref\GRS{R.~Gregory, V.~A.~Rubakov and S.~M.~Sibiryakov,
``Brane worlds: The gravity of escaping matter,''
hep-th/0003109.}
\lref\DRT{S.~L.~Dubovsky, V.~A.~Rubakov and P.~G.~Tinyakov,
``Is the electric charge conserved in brane world?,''
hep-ph/0007179.}
\lref\HoIt{G.~T.~Horowitz and N.~Itzhaki,
``Black holes, shock waves, and causality in the AdS/CFT correspondence,''
hep-th/9901012, {\sl JHEP}  {\bf 9902}, 010 (1999).}
\lref\CoSm{S.~Coleman and L.~Smarr,
``Are There Geon Analogs In Sourceless Gauge - Field Theories?,''
{\sl Commun.\ Math.\ Phys.}  {\bf 56}, 1 (1977).}
\lref\CHR{A. Chamblin, S.W. Hawking, and H.S. Reall, ``Brane-world black
holes,'' hep-th/9909205, {\sl Phys.\ Rev.}  {\bf D61}, 065007 (2000).}
\lref\AIMVV{I.~Y.~Aref'eva, M.~G.~Ivanov, 
W.~Muck, K.~S.~Viswanathan and I.~V.~Volovich,
``Consistent linearized gravity in brane backgrounds,''
hep-th/0004114.}
\Title{\vbox{\baselineskip12pt
\hbox{hep-th/0009176}\hbox{MIT-CTP-3024}
}}
{\vbox{\centerline{Effective theories and black hole production}
\vskip2pt\centerline{in warped compactifications}
}}
\centerline{{\ticp Steven B. Giddings}\footnote{$^\dagger$}
{Email address:
giddings@physics.ucsb.edu} }
\bigskip\centerline{ {\sl Department of Physics}}
\centerline{\sl University of California}
\centerline{\sl Santa Barbara, CA 93106-9530}
\bigskip
\centerline{{\ticp Emanuel Katz}\footnote{$^*$}{Email
address: amikatz@mit.edu }}
\bigskip\centerline{{\sl Center for Theoretical Physics}}
\centerline{\sl Massachusetts Institute of Technology}
\centerline{\sl Cambridge, MA 02139}
\bigskip
\centerline{\bf Abstract}
We investigate aspects of the four-dimensional effective description of
brane world scenarios based on warped compactification on anti-de
Sitter space.  The low-energy dynamics is described by visible matter
gravitationally coupled to a ``dark'' conformal field theory.  We give the
linearized description of the 4d stress tensor corresponding to an
arbitrary 5d matter distribution.  In particular a 5d falling particle
corresponds to a 4d expanding shell, giving a 4d interpretation of a
trajectory that misses a black hole only by moving in the fifth dimension.
Breakdown of the effective description occurs when either five-dimensional
physics or strong gravity becomes important.  In scenarios with a TeV
brane, the latter can happen through production of black holes near the TeV
scale.  This could provide an interesting experimental window on quantum
black hole dynamics.

\Date{}

\newsec{Introduction}

It is an old idea that, as an alternative to compactification, the observed
Universe instead lives on a brane in a higher-dimensional space.  Such
``branification'' scenarios had however until recently been hard to
realize, largely because of the difficulty of recovering four-dimensional
gravitational dynamics.  Two new approaches have changed this and at the
same time suggested new views of the origin of the hierarchy of scales in
physics.  The first, pursued by \refs{\ADD}, is a hybrid of branification
and compactification, in which matter is confined to a brane and then
large-radius compactification of the extra dimensions yields
four-dimensional gravity at long distances.  

A more recent approach utilizes warped
compactifications to achieve effectively four-dimensional gravitational 
dynamics.  A
outline of such a picture has been provided by the RSII model\refs{\RSII}.
This utilizes a ``Planck brane'' that serves as the boundary of
five-dimensional anti-de Sitter space, and the curvature of anti-de Sitter
space effectively ``localizes'' low-energy gravity to the brane.  Related
models are the RSI model\refs{\RSI} in which AdS is terminated above the
horizon by a ``negative tension brane,'' and the model of Lykken and
Randall \refs{\LyRa} in which visible sector matter lives on a probe brane.
None of these are fundamental pictures as they do not provide a microscopic
dynamics for the Planck, ``negative-tension,'' and probe branes, but recent
work in string theory has begun to provide descriptions of such objects.
In particular
\refs{\Hver} has given a geometrical realization of an object akin to a
Planck brane, and \refs{\KlSt,\MaNu} have provided geometrical realizations 
of objects similar to ``negative-tension'' branes.  At the same time, these
models have been connected to renormalization group flows in
four-dimensional gauge theories through the AdS/CFT correspondence.

In providing a new view of the hierarchy problem, either through large
radius or other geometrical mechanisms, these scenarios suggest the
exciting possibility that quantum gravity effects could be observed at
scales far below the usual Planck scale, and perhaps even near the TeV
scale.  They also suggest the possibility of interesting new gravitational
phenomena, particularly in scenarios with infinite extra dimensions ({\it
e.g.} RSII) and with non-trivial curvature and horizon structure of the
resulting spacetime.

Some aspects of this gravitational dynamics has been studied in
\refs{\EHM\EHMii\GaTa-\GKR}.  In particular, \GKR\ studied linearized
gravity in the RSII scenario, and gave both prescriptions for computing
propagators and a general picture of the structure of black holes bound to
the Planck brane.  The latter were found to be pancake-like objects, whose
transverse sizes are logarithmically smaller than their four-dimensional
Schwarzschild radii.  Cosmology of these scenarios has also been
extensively studied (see \eg\ \refs{\Kal\CGKT\CGRT-\KKOP}) 
with suggestions that they offer new approaches
to the cosmological constant problem 
\refs{\VerVer\EVer\ADKS\KSS\KSSii\FLLN\deAli\CEGH\deAlii-\HLZ}.

Many open questions remain, however, in the RSII scenario and its variants.
One set of questions centers on the four-dimensional representation of the
five-dimensional dynamics.  In particular, localization of gravity is
not complete and in the RSII scenario there is a gapless spectrum of
analogs to Kaluza-Klein modes that are weakly coupled to excitations on the
brane.  Therefore a four-dimensional
low-energy effective field theory does not follow from the usual
Kaluza-Klein reasoning, and so one challenge has been to deduce what this
effective theory is.  It has previously been
argued\refs{\Hver,\WittITP,\Gubs,\GKR,\HMS} that the bulk dynamics can be
replaced via the AdS/CFT correspondence by a conformal field theory on the
brane, and this suggests an answer, namely that the effective 
field theory is provided by
conformal field theory coupled to the visible sector solely through
gravity.  This paper amplifies on this statement, clarifies the role of the
cutoff, which in RSII is expected to be at the AdS radius
scale,
and provides one entry in
the map between the five- and four-dimensional descriptions by computing
a linearized approximation to the four-dimensional stress tensor
corresponding to an arbitrary five-dimensional matter distribution.  This
stress tensor is both conserved and traceless.  Corresponding statements
should hold for other warped compactification scenarios, using realizations
of the AdS/CFT correspondence in more general warped compactifications.

Given the novelties of the gravitational dynamics, for
example the above picture of black holes, one is also prodded to
investigate whether this field theory has unusual properties.  For example,
consider the following question\GKR:\foot{This question was asked by L.
Susskind.}  suppose that a particle is launched towards a black hole on the
brane with zero four-dimensional impact parameter, but such that it follows
a trajectory that misses the black hole through the fifth-dimension.  Does
this correspond in the four-dimensional perspective to matter that  enters a
black hole and exits the opposite side?  This would surely be a radical
departure from usual four-dimensional effective theory!

However, standard AdS/CFT reasoning suggests a more mundane answer.  In the
UV/IR correspondence outlined in \refs{\SuWi}, a state deep in AdS
corresponds to a state in the far infrared of the corresponding field
theory.  This suggests that a falling particle corresponds to a state that
spreads.  Indeed, using our results for the stress tensor we find that in
the four dimensional description, the falling particle corresponds to an
expanding shell of CFT matter.  The condition that the five-dimensional
trajectory misses the black hole becomes the four-dimensional 
statement that the
shell misses by expanding to a size larger than the black hole.  

It is important to emphasize that the CFT description is an effective
description, and 
another interesting set of questions therefore regards breakdown of the effective
field theory and the question of whether strong gravitational dynamics -- for
example black hole formation -- is observable at scales far below the
four-dimensional Planck scale.  We investigate the scales at which
scattering experiments would be expected to encounter dynamics beyond the
four-dimensional description in the three scenarios outlined, RSII, the
probe brane scenario, and terminated AdS.  In particular, in the latter
scenario with a certain set of assumptions it appears 
possible to create black holes that decay into
observable matter in scattering experiments in the vicinity of the TeV
scale.  This exciting possibility deserves more theoretical investigation;
in particular through construction of concrete models with the required
properties.  

In outline, section II of this paper discusses conformal field theory as
the 4d low-energy effective theory of RSII. Section III computes the
linearized effective stress tensor of bulk matter, as well as solving a
corresponding simpler problem of the 4d scalar profile of a
five-dimensional scalar source.  It also elaborates on the black hole flyby
scenario mentioned above.  Section IV then discusses questions of the scale
of breakdown for the 4d effective theories, and of the possibility of
low-energy black hole production.  Section V closes with conclusions.

We have been informed that related work in progress\APR\ also addresses
issues of black hole production and corrections to the effective theory in
TeV brane scenarios. 

\newsec{The effective theory of RSII}

We begin with a quick review of the RSII scenario, and of
its transcription into conformal field theory via the AdS/CFT
correspondence\refs{\WittITP,\Gubs,\GKR} in which we will offer some
refinements.  The upshot of this discussion is that the low-energy
effective field theory for the RSII scenario consists of visible 4d matter
gravitationally coupled to dark matter described by a cutoff CFT.
Subsequent sections will explore consequences and extensions of this picture.

The RSII scenario is of course just an example of a much broader class of
warped compactifications, which have recently been widely studied both in
the context of model building, and in the context of string theory and the
correspondence between renormalization group flows and supergravity
geometries.  While many of our comments will be made within the framework
of this greatly simplified example (for which the only known microscopic
construction is \refs{\Hver}), corresponding arguments should apply to
other models including those with stringy realizations.  In particular
later sections will also comment on other variants of the RSII scenario
(those with a terminated AdS space or with a probe or ``TeV'' brane) and
their possible stringy realizations.

We therefore
begin by considering the geometry with a single ``Planck'' brane.  Although
our central interest is dimension $d=4$, most of the relevant formulas
easily generalize and will be given in arbitrary dimension.  We assume that
matter fields, denoted by $\psi$, live only on this Planck brane.  
The action
is 
\eqn\dact{
S = \int{d^{d+1}}X\sqrt{-G}({M^{d-1}}\calR-\Lambda) + 
\int{d^{d}{x}}
\sqrt{-\gamma}\left[{\cal L}(\gamma,\psi) -\tau\right]
\ }
where $G$, $M$, $\calR$, and $\Lambda$ are the $d+1$ dimensional metric, 
Planck mass, curvature
scalar, and cosmological constant respectively, 
$\gamma$ is the induced metric on the Planck brane,  
${\cal L}$ is the action of matter on the brane, and $\tau$ is the brane
tension.
The bulk AdS metric is
\eqn\bmet{ds^2= {R^2\over z^2}\left(dz^2 + dx_d^2\right) }
in $d+1$-dimensional coordinates $X=(x,z)$; 
here $dx_d^2$ is the $d$-dimensional Minkowski metric and 
the AdS radius $R$ is given by 
\eqn\adsrad{
{\Rads} = \sqrt{-d(d-1)M^{d-1}\over{\Lambda}}.}
The brane tension is fine tuned to the value
\eqn\tens{
\tau={4(d-1){M^{d-1}}\over{\Rads}}}
in order to maintain a Poincar\'e invariant Planck brane.  We may take the
Planck brane to reside at an arbitrary elevation $z=\rho$.

As argued in \refs{\RSI,\RSII,\GKR}, at long distances compared to $R$, the
gravitational dynamics appears $d$-dimensional.  However, there is also a
gapless spectrum of weakly-coupled bulk modes.  An obvious question is what
serves as a $d$-dimensional low-energy effective field theory describing
the dynamics.  Within string theory, an answer to this is provided by the
conjectured AdS/CFT correspondence \refs{\WittITP,\Gubs,\GKR}.

To see this, recall that the AdS/CFT correspondence equates the
$d+1$-dimensional bulk gravity (or more precisely, string theory) 
functional integral to a generating function
in the CFT.  A regulator is provided by excluding the AdS volume
outside $z=\rho$.  Suppose that we put the fluctuating 
metric in a gauge such that near
this boundary 
\eqn\asympmet{ds^2 = {R^2\over z^2}\left[dz^2 + g_{\mu\nu}(z,x)dx^\mu
dx^\nu\right]\ .}
The induced metric $\gamma$ on the boundary $z=\rho$ is thus
\eqn\indmet{ds_d^2 =
{R^2\over \rho^2} g_{\mu\nu}(\rho,x)dx^\mu dx^\nu\equiv
\gamma_{\mu\nu} dx^\mu dx^\nu\ .}
Define the functional integral over bulk metrics $G$ 
for fixed boundary metric $\gamma$ as 
\eqn\zdef{Z[\gamma,\rho]=
\int_{\gamma}  \cald G e^{i\int d^{d+1}X \sqrtG (M^{d-1} \calR-\Lambda)
 +2iM^{d-1} \int d^dx
\sqrt{-\gamma} K}}
where $K$ is the extrinsic curvature of the boundary.  The AdS/CFT
correspondence then states that for small fluctuations  about the flat
boundary geometry, $g_{\mu\nu}= \eta_{\mu\nu} + h_{\mu\nu}$
\eqn\adstwo{  \lim_{\rho\rightarrow0} e^{-iS_{\rm grav}[\gamma] }
 Z[\gamma,\rho]  = 
\biggl\langle e^{i\int
h_{\mu\nu}T^{\mu\nu} } \biggr\rangle_{\rm CFT}\ .}
Here $S_{\rm grav}$ is a counterterm action formed purely from the induced
metric $\gamma$ \refs{\HeSk,\BaKr}; in the case $d=4$
\eqn\sct{S_{\rm grav}=\int d^4 x \sqrt{-\gamma}\left[{6M^3\over R} +{RM^3\over 2}
\calR(\gamma) -2M^3R^3 \log(\rho) \calR_2(\gamma)\right] }
where
\eqn\rsqdef{\calR_2 = -{1\over 8} \calR_{\mu\nu} \calR^{\mu\nu} + {1\over
24} \calR^2\ .}

While the AdS/CFT correspondence was originally stated in terms of small
fluctuations, a natural assumption is that it extends to more general
boundary geometries.  We therefore assume that the CFT generating
functional can be written as a functional integral over the CFT degrees of
freedom, which we collectively denote $\chi$, in the background metric
$g_{\mu\nu}$, and that the correspondence thus becomes
\eqn\fctlint{  \lim_{\rho\rightarrow0} e^{-iS_{\rm grav}[\gamma] }
 Z[\gamma,\rho] =\int \cald\chi e^{i\int d^d x \sqrt{-g} \calL_{\rm
 CFT}(g_{\mu\nu},\chi)} \ .}
Following the ideas of the UV/IR correspondence\refs{\SuWi}, we
connect this with the RS scenario by extending the conjecture
to a statement with a finite cutoff, and assume that
\eqn\finads{e^{-i \Sct[\gamma]} Z[\gamma,\rho] = \int [\cald\chi]_\rho 
e^{i\int d^d x \sqrt{-g} \calL_{\rm
 CFT}(g_{\mu\nu},\chi)} \ }
where the on the RHS $\rho$ provides the cutoff scale for the CFT.  
While a precise
description of this cutoff in the language of the CFT is not known,
for sake of intuition one
may imagine that it is for example given by only considering 
fluctuations on scales $\Delta x$ such that 
\eqn\cutdef{g_{\mu\nu} \Delta x^\mu \Delta x^\nu > \rho^2\ .}
In particular, notice that since the only dependence of the CFT on the scale of the
metric is through the cutoff, this implies
\eqn\cutrel{\int [\cald\chi]_\rho 
e^{i\int d^d x \sqrt{-g} \calL_{\rm
 CFT}(g_{\mu\nu},\chi)} = \int [\cald\chi]_R 
e^{i\int d^d x \sqrt{-\gamma} \calL_{\rm
 CFT}(\gamma_{\mu\nu},\chi)}\ }
where on the RHS the cutoff is thought of as restricting to fluctuations
with
\eqn\newcut{\gamma_{\mu\nu} \Delta x^\mu \Delta x^\nu > R^2\ .}

From \finads\ and \cutrel\ we therefore see that the integral over the bulk
modes can be replaced by a correlator in the CFT, as originally proposed in
\refs{\WittITP,\Gubs,\GKR}, with a cutoff given by $R$.  Specifically, 
$d$-dimensional dynamics is summarized by a functional integral of the form
\eqn\deft{\int [\cald \gamma\cald \psi\cald\chi]_R 
e^{i\int d^dx\sqrt{-\gamma}
\left[{1\over2}
\calL(\gamma,\psi) + \calL_{\rm CFT}(\gamma,\chi) + \calL_{\rm grav}(\gamma)
-\tau\right]}(\cdots)\ .}
For consistency with the cutoff \newcut\ the other modes also presumably
should have a corresponding cutoff, as indicated.  One consistency check on
this approach is cancellation of the brane tension $\tau$ by the
corresponding term in $\Sct$, using \tens.  This indicates that the
low-energy effective field theory for the system, up to the scale
determined by $R$, is the theory of brane-matter gravitationally coupled to
``dark'' matter described by the CFT.  The $d$-dimensional Planck mass
follows from the $d$-dimensional version of \sct, and is given by
\eqn\fdpm{M_d^{d-2}= {RM^{d-1}\over d-2} \ .}

\newsec{Effective stress tensor of bulk matter}

We now investigate some of the consequences of the above identification of
the CFT as the low-energy effective field theory for the RSII scenario.  In
particular, we start by giving an entry in the bulk to boundary dictionary,
by computing a linearized approximation to the CFT stress tensor
corresponding to a perturbation in the bulk.  We then investigate the
particular case of a particle freely falling into the bulk.  

Using this calculation, we discuss a test of the AdS/CFT
correspondence and of our effective description of RSII:
suppose that we shoot a particle towards a black hole with zero 4d
impact parameter, but such that it will miss the black hole 
through the $z$ direction.
How does a 4d observer understand the failure of the black hole to absorb
the particle?

\subsec{General results}

In this subsection we turn to the problem of deriving the $d$-dimensional
brane stress tensor that corresponds to a general $d+1$-dimensional bulk
matter distribution.  In general this is a difficult problem, requiring
solution of the bulk Einstein equations, so we will only give a linear
treatment.

The basic strategy is as follows.  Ref.~\refs{\GKR}
computes the linearized bulk gravitational field of a general matter
perturbation.  This in particular gives the linearized metric and therefore
Einstein tensor induced on the brane.  We can then read off the matter stress
tensor from the right hand side of the $d$-dimensional Einstein equations
along the brane.

Although the resulting stress tensor has a number of special properties,
we have not yet found a particularly illuminating  expression for it.
However, in the next subsection we specialize to the case of a particle
falling into the bulk; in the long distance approximation the corresponding
stress tensor simplifies substantially.

In studying gravitational perturbations it proves convenient to introduce 
the proper ``height''
coordinate $y$, given by
\eqn\yzreln{z=Re^{y/R}\ }
in terms of which the linearized metric takes the form
\eqn\linmet{ds^2 = dy^2 + e^{-2y/R}(\eta_{\mu\nu} + h_{\mu\nu})dx^\mu
dx^\nu\ .}  Eqs.~(3.20), (3.24), and (3.26) of \GKR\ then give  
the linearized
bulk Einstein equations in terms of the metric perturbation ${\bar
h}_{\mu\nu}= h_{\mu\nu}-\hf h \eta_{\mu\nu}$ as:
\eqn\dhdy{
\partial_y\left( e^{-2y/\Rads}\partial_y h\right) = 
{1\over (d-1)M^{d-1}}  \biggl[ T_{\mu}^{\mu} - (d-2)e^{-2 y/R}\, T^y_y\biggr] \ ,
}
\eqn\muysol{
\partial_y\partial^{\nu} h_{\mu\nu} = \partial_y\partial_{\mu}
h + {T_{\mu}^y\over M^{d-1}}\ ,
}
and
\eqn\munueq{
   \eqalign {
\sq{ \bar{h}_{\mu\nu} }
&= { \eta_{\mu\nu}\over 2 }\, e^{yd/R} \, \partial_y
(e^{-yd/R}\partial_y{h})\cr
&+ e^{2y/\Rads}
(-\eta_{\mu\nu}\partial^{\lambda}\partial^{\sigma}{\bar{h}_{\lambda\sigma}} +
\partial^{\lambda}\partial_{\mu}{\bar{h}}_{\nu\lambda}
+\partial^{\lambda}\partial_{\nu}{\bar{h}}_{\mu\lambda} )\cr
&- { e^{2y/R}\over M^{d-1}} \, T_{\mu\nu}\ . \cr}
}

The right hand side of \munueq\ is determined by the stress tensor and the
solutions of eqs.~\dhdy, \muysol.  This equation can then be solved for
$h_{\mu\nu}$ using the scalar Neumann Green function $\Delta_{d+1}$,
satisfying 
\eqn\branegf{\eqalign{
\sq \Delta_{d+1} (X,X^{\prime}) &= {\delta^{d+1}(X - X^{\prime})\over {\sqrt {-
G}}}\ ,\cr
\partial_y \Delta_{d+1}(X,X')\vert_{y=0}&=0\ , }
}
and which was derived in \GKR.  In the present situation we need the
retarded propagator rather than the Feynman propagator; the relation
between these and approximate expressions for them are given in the appendix.
The resulting expression for the metric
has three terms arising from the three lines of
\munueq.  
However, the second term 
is inessential as a short calculation shows it to be pure gauge on the
brane. 
Therefore its contribution drops when we
compute the Einstein tensor on the brane.  

One must also specify boundary
conditions at the brane; in the case of a surface stress tensor 
\eqn\braneT{
T_{\mu\nu}^{\rm brane} = S_{\mu\nu}(x)\delta(y)\, , \hskip.2in T_{yy}^{\rm
brane} = T_{y\mu}^{\rm brane} = 0}
these become 
\eqn\bcond{
\partial_y(h_{\mu\nu} - \eta_{\mu\nu}h )_{\vert y=0} = - {S_{\mu\nu}(x)\over
2M^{d-1} }\, .}

In order to simplify the resulting expression for the metric, it is useful
to rewrite the scalar Green function 
in terms of a new function $F$ as 
\eqn\gammadef{\Delta_{d+1}(y,x;y',x') = e^{(d-2)y'/R} \partial_{y'} 
\left[ e^{-(d-2)y'/R}
F_{y'}(y;x-x')\right] \ .} 
One nice property of this redefinition is immediate:  one readily checks
that 
\eqn\fdgf{ \int_0^\infty dy' e^{(2-d)y'/R} \Delta_{d+1}(X,X') = -
F_0(y;x-x')}
satisfies the equation for the $d$-dimensional Green function, and so
\eqn\gammagf{F_0(y;x-x')= -\Delta_d(x,x')\ .}
Using this and integrating 
by parts gives the contribution to $\hbmn$ from the first line in 
\munueq\ as 
\eqn\honedef{{\bar h_{\mu\nu}^{(1)}}(x,y) = {\eta_{\mu\nu} \over 2(d-1) M^{d-1}}
\int dV' \partial_{y'}F_{y'}(y;x-x')\left[e^{2y/R} T_\mu^\mu(X') - (d-2)
T_y^y(X')\right]\
.}

Eq.~\honedef\ combines with the terms induced by the stress tensor and the
surface stress \braneT\ to give a complete expression of the form
\eqn\fullmet{\eqalign {\hbmn &= -{1\over M^{d-1}} 
\int dV' e^{dy'/R} \partial_{y'} 
\left[ e^{-(d-2)y'/R}F_{y'}(y;x-x')\right] T_{\mu\nu}(X')\cr 
&+{\eta_{\mu\nu} \over 2(d-1) M^{d-1}}
\int dV' \partial_{y'}F_{y'}(y;x-x')\left[e^{2y/R} T_\mu^\mu(X') - (d-2)
T_y^y(X')
\right]\cr
&+{\bar h}^S_{\mu\nu}+ {\bar h}^{\rm gauge}_{\mu\nu}\ .}}
Here ${\bar h}^{\rm gauge}_{\mu\nu}$ is the piece that is pure gauge on the
brane, mentioned above,
and ${\bar h}^S_{\mu\nu}$ is the contribution due to the surface stress (we
will see an example of this shortly).

For simplicity consider a purely bulk distribution ($S_{\mu\nu}=0$).  The
four-dimensional effective stress tensor is readily computed from \fullmet\
via Einstein's equations,
\eqn\efstress{T^{\rm eff}_{\mu\nu} = 2M_d^{d-2} {}^{(d)}{\cal
G}^{\mu}_{\nu} = - { R M^{d-1}\over d- 2} (\partial^2 \hbmn + \etamn
\partial^\alpha\partial^\beta {\bar h}_{\alpha\beta} - \partial_\mu
\partial^\alpha {\bar h}_{\alpha\nu} - \partial_\nu\partial^\alpha {\bar
h}_{\alpha\mu} )\vert_{y=0}\ ,}
with $\hbmn$ given by \fullmet.  The contribution of 
${\bar h}^{\rm gauge}_{\mu\nu}$ drops out.

One may expand out the expression \efstress\ to write it explicitly in
terms of the bulk stress tensor $T_{\mu\nu}$:
\eqn\bulkstress{\eqalign{T^{\rm eff}_{\mu\nu} = {R\over d-2} \int dV' 
\Biggl\{&
e^{dy'/R} \partial_{y'} \left[ e^{-(d-2)y'/R} F_{y'}(0;x-x')\right]
( \partial^2 T_{\mu\nu} + \eta_{\mu\nu} \partial^\alpha\partial^\beta
T_{\alpha\beta}  \cr & - \partial^\alpha \partial_\mu T_{\alpha\nu} -
\partial^\alpha \partial_\nu T_{\alpha\mu} ) \cr &
+{1\over d-1} \partial_{y'}
F_{y'}(0;x-x') (\partial_\mu\partial_\nu -\eta_{\mu\nu} \partial^2)
\left[e^{2y'/R} T_\mu^\mu - (d-2) T_y^y \right] \Biggr\}\ .}}
Note that $T^{\rm eff}_{\mu\nu}$ satisfies
two important properties.  First, from its construction and the Bianchi
identities, it is conserved:
\eqn\tcons{ \partial^\mu T^{\rm eff}_{\mu\nu} =0\ .}
Secondly, one may readily verify that it is traceless,
\eqn\ttrac{ \eta^{\mu\nu} T^{\rm eff}_{\mu\nu} =0\ ,}
which accords nicely with its interpretation as arising from a conformal
field theory on the brane.  Indeed, this easily follows from the $(yy)$
Einstein equation, which states ({\it cf.} \GKR\ eq. (3.14))
\eqn\Gyy{{}^{(d)}{\cal R} + {(d-1)\over R} \partial_y h \epsilon(y) =
-{T^y_y \over M^{d-1}}\ }
where $\epsilon(y)$ is a step function.  On the brane $\partial_y h$ and
$T_y^y$ vanish (the former by \bcond), implying ${}^{(d)}{\cal R}(y=0)=0$, and thus
$T^{\rm eff}=0$.

Note that in the above discussion we have said nothing about bending of the
brane, which was described in \refs{\GaTa,\GKR}.  The reason for this is
that we are interested in the metric on the brane, and for this it is best
to work in a gauge where the brane is straight.  In \GKR\ the resulting
metric was computed by first working in the bent gauge, and then
transforming back, but an equivalent result is found by working directly in
the straight gauge.\foot{For purposes of measurements on the brane, the
apparent breakdown of the linearized approximation at $y\rightarrow\infty$
may be ignored; for another treatment of these matters see \refs{\AIMVV}.}

\subsec{Matter on the brane}

In order to illustrate this equivalence -- and because the result will be
used in the next subsection -- we'll compute the linearized metric and
effective stress tensor due to matter on the brane in this approach.
Specifically, suppose that there is a surface stress of the form \braneT,
but that otherwise $T_{IJ}=0$.  The field eqs.~\dhdy-\munueq\ should then
be solved subject to the boundary conditions \bcond.  By tracing the latter
can
be rewritten in terms of $\bar h$, and take the form
\eqn\hbbc{ \partial_y {\bar h}_{\mu\nu}\vert_{y=0}= -{1\over
2M^{d-1}}\left[S_{\mu\nu} - {\eta_{\mu\nu}\over 2(d-1)} S\right]\ .}
By Green's theorem these give a contribution
\eqn\hbS{ {\bar h}^S_{\mu\nu}(X) = -{1\over2M^{d-1}} \int d^dx'
\Delta_{d+1}(X;0,x') \left[S_{\mu\nu}(x') - {\eta_{\mu\nu}\over 2(d-1)}
S(x')\right]}
to the metric.  As above, the second term on the RHS of \munueq\ is pure
gauge, and the third term vanishes, so the remaining contribution comes
from the first term.  The trace equation \dhdy\ and the boundary condition
\bcond\ imply 
\eqn\heqn{\partial_y h = {e^{2y/R} \over 2(d-1) M^{d-1} } S\ ,}
which gives a contribution
\eqn\Str{ {\bar h}^{(1)}_{\mu\nu} = {\eta_{\mu\nu}(2-d)\over 4(d-1) R
M^{d-1} } \int d^dx' S(x') \int dy' e^{(2-d)y/R} \Delta_{d+1}(X,X')\ .}
The integral over $y'$ is  eliminated by using the identities \fdgf\ and
\gammagf, and the combined expressions \hbS\ and \Str\ yield
\eqn\branesol{
   \eqalign {
\bar{h}_{\mu\nu} (x) &= -{1\over 2M^{d-1}} \, \
\int \, d^d x^{\prime} \Biggl\{
\, \Delta_{d+1} (x,0;x^{\prime},0) S_{\mu\nu}(x') - \cr
& \eta_{\mu\nu} \biggl[ \Delta_{d+1} 
(x,0;x',0) - {(d-2) \over \Rads} \Delta_{d}(x,x')\biggr]  \,
{S^{\lambda}_{\lambda}(x') \over 2 (d-1)}\Biggr\}\ 
}}
in agreement with \GKR.  In particular, this expression may be evaluated
for a stress tensor corresponding to a point mass at rest on the brane at
${\vec x}=0$,\foot{The extra factor of two is present because of the
orbifold boundary conditions, and compensates the 1/2 in \deft.} 
\eqn\istress{T_{tt} = 2m \delta^{d-1}(x)\delta(y)\ .}
Using the long-distance expansion of the propagator\GKR, 
\eqn\propexp{\Delta_{d+1}(x,0;x',0)={d-2\over R}
\Delta_d(x,x')\left[1+\left({R^{d-2}\over r^{d-2}}\right)\right]\ ,}  
this gives the $d=4$ expression
\eqn\ptmet{ {\bar h}_{tt} = {m\over 2\pi R M^3 r} \left[1+
\calo\left({R^2\over r^2}\right)\right]\ ,\ {\bar h}_{ij} =
\calo\left({mR\over M^3r^3}\right)\ .}

\subsec{The falling particle}

The above  
expression \bulkstress\ for the stress tensor appears rather complicated, but
simplifies significantly in the long-distance limit.  To illustrate this,
we compute the effective stress tensor of a particle falling into the
bulk. (The corresponding approximate 
metric has also been computed by Gregory, Rubakov, and
Sibiryakov in \refs{\GRS}.)  This case will also apply to our later
discussion of black hole flybys; by performing a boost along the brane we
get a trajectory that can sail behind a black
hole through the extra dimension.  

Concretely, consider a particle of mass $m$ that is stuck to the brane at
${\vec x}=0$ until time $t=0$ and then released and allowed to freely fall
into the bulk.  The trajectory for $t>0$ is easily seen to be given by the
equation 
\eqn\traj{z^2-t^2 = R^2\ .}
For $t<0$ the only nonzero component of the stress tensor is given by \istress.
%
%
For $t>0$ the stress tensor is given by the general formula 
\eqn\gstress{T_{IJ}= m {dX_I\over d\tau} {dX_J\over dt}
{\delta^{d-1}(x-x(t))\delta(y-y(t))\over \sqrt{-G}}\ ,}
which in the present case gives nonvanishing components
\eqn\Ttt{T_{tt}= m  e^{(d-2)y/R} \delta^{d-1}({\vec
x})\delta(y-y(t))\ ,}
\eqn\Tyy{T_{yy}=m {t^2\over R^2} e^{(d-2)y/R} \delta^{d-1}({\vec
x})\delta(y-y(t))\ ,}
and
\eqn\Tty{T_{ty}=-m {t\over R} e^{(d-2)y/R} \delta^{d-1}({\vec
x})\delta(y-y(t))\ .}

Therefore the contribution to the metric from the trajectory for $t<0$
is a special case of the general surface-stress result of the preceding
subsection, \branesol, with 
\eqn\sttpt{S_{tt}=2m\delta^{d-1}(x)\theta(-t)\ ,\ S_{\mu i}=0\ .}
The contribution to the metric from the second half of the trajectory is
given by our formula \fullmet.  Specifically, rewriting the $t>0$ stress tensor
as
\eqn\fallstress{ T_{IJ} = \calS_{IJ}(t,y)\delta(y-y(t))\ ,}
we find
\eqn\fmet{\eqalign{\barh^>_{\mu\nu}(x) = -{1\over M^{d-1}}\int_{t'>0}& d^dx'\Bigl\{
\partial_y\left[e^{-(d-2)y/R}
F_y(0;x-x')\right]\calS_{\mu\nu}(x',y)\cr & -{\etamn\over
2(d-1)} e^{-(d-2)y/R} \partial_yF\left[\calS^\mu_\mu -(d-2)e^{-2y/R}
\calS^y_y\right]\Bigr\}\vert_{y=y(t)}\cr & + {\bar h}^{\rm gauge}_{\mu\nu}\ .}}

The expression for the effective stress tensor follows directly from
computing the Einstein tensor \efstress\ from these expressions for the
linearized metric.
In order to gain some intuition
for this expression, consider the approximation of 
distances and times much greater than the AdS scale $R$, which we've seen
is the cutoff for the effective theory: 
\eqn\smallR{x^2-t^2\gg R^2\ .} 
In this limit the Green function simplifies dramatically (see appendix),
\eqn\appgf{ F_y(0;x-x')\simeq {1\over 2\pi}\delta(z^2+(x-x')^2)\theta(t-t')\ ,}
and the trajectory \traj\ becomes
\eqn\apptraj{ z=Re^{y/R}\simeq t\ .}

Defining $r=|x|$, in $d=4$ the approximate metric is then
\eqn\apim{ \barh_{\mu\nu}\simeq {m\over 2\pi M^3 R r}
\delta^t_\mu\delta^t_\nu}
for $r>t$, and
\eqn\apfm{ \barh_{\mu\nu} \simeq {m\over 2\pi M^3 R}\left[\left({3\over
2t}-{{\vec x}^2\over 2t^3}\right)\delta^t_\mu\delta^t_\nu + 
{t^2-{\vec x}^2\over
4t^3}\etamn\right]}
for $r<t$, as in \GRS.  A straightforward computation shows that the
Einstein tensor of both of these metrics vanishes!  Thus the effective
stress tensor is concentrated on the surface where they match, $r=t$.
This stress tensor is
\eqn\apT{T^{\rm eff}_{\mu\nu} \simeq {m\over 4\pi t^2} \delta(t-r) u^\mu
u^\nu\ }
where $u^\mu= x^\mu/t$.

The effective stress tensor of the conformal field theory configuration
describing a falling particle is 
thus concentrated on a thin shell of radius $r$ which expands
outward with time, $r=t$.  We can estimate the thickness of the shell by
investigating the size of the leading corrections in the limit \smallR.
One readily sees that the metric is corrected at order $R^2/t^2$,
$R^2/r^2$,
both due
to corrections to the trajectory and to the Green function.  This suggests
that the thickness of the shell of CFT matter is the expected ${\cal
O}(R)$, the cutoff length scale.

This spreading behavior appears to be quite generic, as one might expect
from the IR/UV duality of the AdS/CFT correspondence.  Another example of
this behavior is
provided by a falling charged particle coupled to a bulk gauge field, as
investigated in \refs{\DRT}.
Indeed, an even simpler example is provided
by a falling particle coupled to a bulk scalar field.  Specifically,
consider a Lagrangian 
\eqn\scalag{ S=-\int dV {1\over 2} (\nabla \phi)^2 -q\int d\tau
\phi(X(\tau)) }
with a coupling of a bulk scalar field $\phi$ 
to a particle of scalar charge $q$ falling along a
trajectory $X(\tau)$.  This determines the scalar field,
\eqn\scalsoln{\phi(X)= q\int d\tau \Delta_{d+1}(X,X(\tau))\ .}
If we assume that the particle again follows the trajectory \traj\ and work
at large distances as compared to $R$ and with $d=4$, then the field at $y=0$ takes the
approximate form
\eqn\largRsol{ \phi(x,t) \simeq -{q\over 2\pi R}\left[{1\over r}
\theta(r-t) + {4R^2 t\over (r^2-t^2)^2}\theta(t-r)\right]\left[1+
\calo\left({R\over r}\right)^2\right]\ .}
If we compute the effective source, $J=\sq_4\phi$, we find it vanishes
except at $r=t$.
Again, subleading $\calo(R)$ corrections appear to smooth this into a shell
of thickness $R$.

Note that similar behavior was found by Horowitz and Itzhaki \refs{\HoIt},
who investigated the CFT stress tensor
corresponding to a particle moving geodesically in the full, infinite AdS.
This work also found a shell expanding outward at the speed of light.
Indeed, the two calculations are directly related in the infrared limit, as
discussed in appendix B.

This behavior can can also be understood directly
in terms of the CFT using an argument due to Coleman and Smarr
\refs{\CoSm}, which shows that a stress tensor that is conserved,
traceless, and has positive energy density will be localized on the light
cone.  The basic idea for the proof is to show that the average squared
energy
radius,
\eqn\ravsq{{\bar r}(t)^2 = {\int d^3x r^2 T_{00} \over  \int d^3x T_{00}} }
satisfies
\eqn\radeq{{d^2 {\bar r}^2 \over dt^2} =2}
from which it immediately follows that a configuration initially localized
at a point will expand on the light cone.
Ref.~\HoIt\ argues that the argument extends even to the quantum
case, where the energy density may be negative, as long as the total energy is
positive.  

These results neglect the backreaction of gravity on the outgoing shell.
It would be interesting to understand what dynamics
results when strong self gravitation of the shell is included.

Vanishing of the Einstein tensor for the metric \apfm\ at first sight leads
to another puzzle.  Specifically, suppose we consider a ``bounce''
trajectory, where the particle follows the trajectory \traj\ for {\it all}
time.  The calculation of the metric above is modified by extending to the
trajectory for $t<0$, but still yields a stress tensor that vanishes
everywhere.  This contradicts our expectation of a shell that collapses and then
reexpands.  However, note that this computation is not complete.  The $z$
coordinates only cover the region outside the AdS horizon, and thus this
calculation would miss the contribution of the piece of the trajectory
behind the past horizon.  If this is not included, energy-momentum
conservation is violated at the horizon, and consequently gravity cannot be
consistently coupled.  A correct calculation includes this piece, but also
requires more information about the structure of the Green function.
Specifically, one needs to know what boundary conditions it obeys in the
far past.  In order to make predictions in the RSII scenario,
one needs to understand the physics determining the boundary conditions at
the past horizon.  Correspondingly, in CFT language one needs to know in
what state the CFT sector began.

\subsec{Black hole flybys}

We now have the necessary tools to discuss particles flying past black
holes through the bulk; for simplicity we discuss the four-dimensional
case.  Specifically, suppose that there is a black hole of mass $m$ located
at ${\vec x}=0$ and that a particle is shot at it with zero four-dimensional
impact parameter, but is allowed to fall in $z$ long enough
that it misses the black hole by passing it in the $z$ direction (see
fig.~1).  
How does a four
dimensional observer describe such an experiment, and in particular does
one see radical departures from usual gravitational dynamics, such as
matter entering and then escaping a 4d event horizon?

\ifig{\Fig\One}{A particle trajectory that misses a black hole on the brane
because of its motion in the extra dimension.}{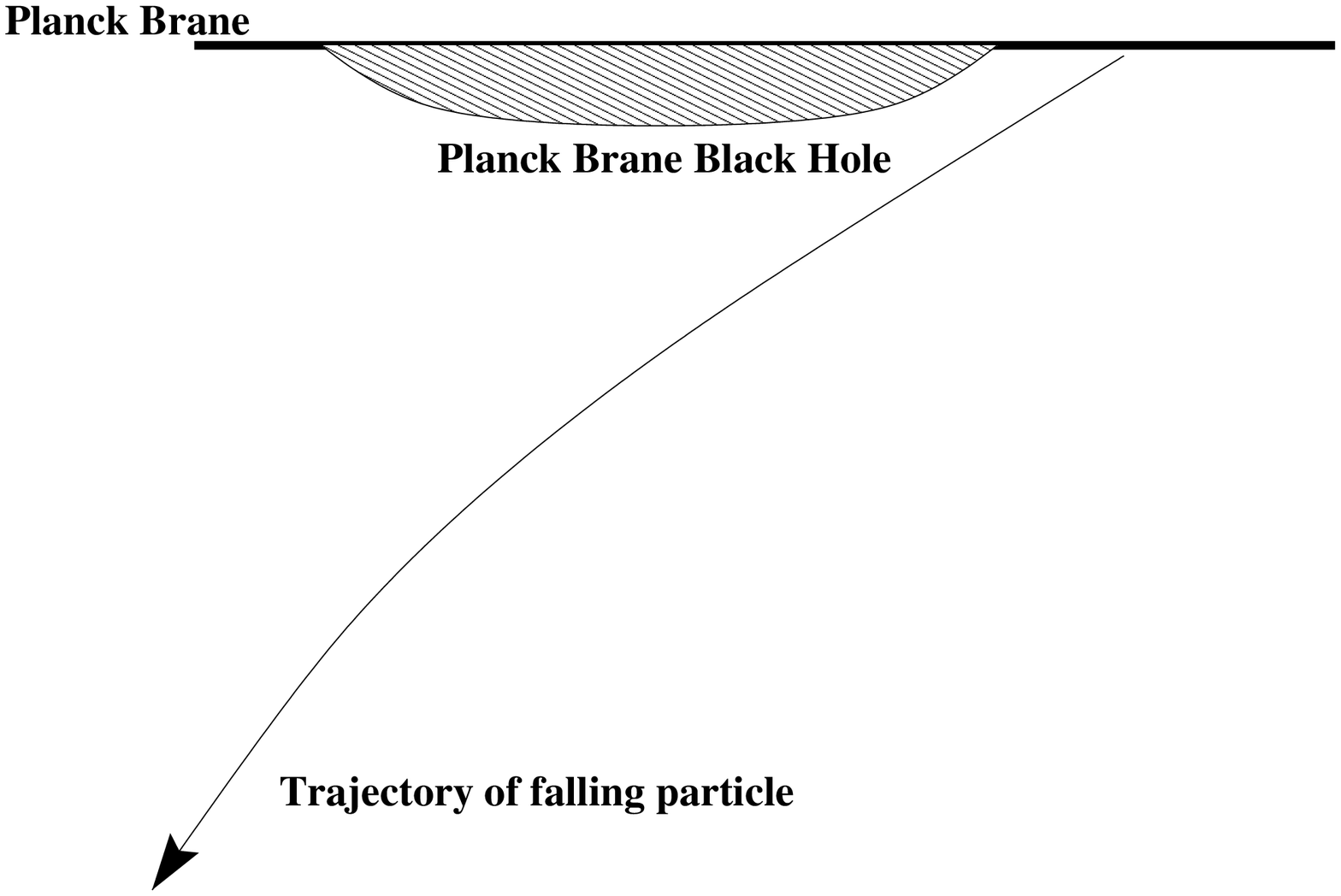}{2.0}

The answer to the latter question is, of course, no.  Indeed, a black hole
with mass and 4d Schwarzschild radius $m$ has a horizon extending to
$z_h\sim m$ in the bulk picture.  In order for the particle to miss the
black hole, the
particle 
must have $z\gg m$ when ${\vec x}=0$.  As we've seen above, in the CFT
description the particle corresponds to a shell of CFT matter.  If it has
reached $z\gg m$ by the time it reaches the black hole at ${\vec x}=0$, then
the shell has expanded to size $r\gg m$ by the time it has reached the
black hole, and is continuing to expand outward.  No novel physics need be
invoked to explain why the shell is not absorbed by the black
hole.\foot{Note, however, that a small piece of the shell may be absorbed
by the black hole; a quantum treatment of the bulk should yield a
corresponding result.}  The process has a perfectly adequate four-dimensional
effective description in terms of the matter conformal field theory coupled
to  four-dimensional gravity.

\newsec{Breakdown of EFT; cutoffs, strong gravity, and black hole production}

Section two argued that at low energies the RS scenario is equivalent to
coupling ordinary matter to a hidden CFT.  Section three provided
illustrations of this statement. 
An obvious question regards the
limitations of this description.  At what scale does it fail?  Is there any
practical advantage or consequence of the five-dimensional description?
And what conclusions can one draw about strong gravitational phenomena,
such as production of black holes in high-energy scattering?

In this section we will first consider the scenario with a single Planck
brane, and then comment on extensions of the discussion to scenarios with
an added probe or ``TeV'' brane or with AdS terminated by a brane-like
object at large $z$ (like the ``negative tension'' brane proposal of \RSI).

\subsec{Scattering on the Planck brane}

The four-dimensional effective action for the scenario with a single Planck
brane is given in \deft.  Recall
that the fundamental parameters determine the 
4d Planck mass by
the relation \fdpm.  We would like to understand what this scenario
predicts for high-energy scattering.

The simplest assumption (if one is not trying to solve the hierarchy
problem) is that the five-dimensional Planck scale and inverse AdS radius,
and hence the four-dimensional Planck scale, are all comparable:
\eqn\mreln{M\sim 1/R\sim M_4 \ .}
In that case all new physics is
clearly at the Planck scale.  How much can this statement be relaxed?  One would
expect observable deviations in microgravity experiments -- as in the
scenario of \refs{\ADD} -- for $R\sim 1mm$.  This puts a lower bound of
$M\roughly> 10^{8} GeV$ on the five-dimensional Planck scale.

Consider now high energy scattering of particles confined to the brane.  From the
bulk perspective, we see that at
distances $\ll R$ the dynamics is effectively five-dimensional.  This is
mirrored in the four-dimensional description of \deft; energies above $1/R$
are past the cutoff and the cutoff CFT description is incomplete.  

Does this mean that we can see what a 4d observer would interpret as strong
gravitational phenomena at energies just above $1/R$?  Clearly not, except
when the parameters satisfy $1/R\sim M$, in which case we are at Planckian
4d energies anyway.
Consider for example black hole production.
There are two types of black holes that one might produce.  The first type
is the AdS/Schwarzschild black hole, which moves freely in the bulk, and in
general will fall towards the AdS horizon once produced.  The threshold
for producing such black holes is the 5d Planck energy $M\roughly>10^{8}
GeV$.  The second type of black hole is bound to the brane, as described in
\refs{\CHR,\GKR}.  The horizon radius of such a black hole is $r_h\sim
m/M_4^2$; this should be larger than the 5d Planck length which implies
 $m>RM^2$.  Since
$RM\roughly>1$, the threshold is again at $M$ or above.  

From this discussion we see that scattering pushes beyond the cutoff scale
at the threshold $1/R$ and in the bulk perspective begins to explore the
extra dimension.  While this may have visible consequences through
production of  the Kaluza-Klein modes, 
strong
gravitational dynamics such as black hole production has a much higher
threshold of $M$, which in the most
``optimistic'' scenario of $M\sim 10^{8} GeV$
is still a long ways off.  

\subsec{Scattering on a probe brane}

The preceding Planck-brane scenario is not favored from the
viewpoint of low-energy phenomenology in any case, given the expected
relation 
\mreln\ between scales.  
Scenarios which try to generate the hierarchy
via the exponential warp factor show more promise.   
Consider first the probe brane scenario of
\refs{\LyRa}.  Here 4d matter is taken to reside on a ``TeV'' brane at an
elevation $z=\rho_T$; the Planck brane is again at $z=\rho$.  
This brane must be stabilized by a mechanism such as
in \refs{\GoWi,\DFGK}.  The 4d Planck mass is again given by \fdpm, but matter on
the TeV brane has its energy redshifted by $\rho/\rho_T$ relative to the
Planck brane.  If $\rho/\rho_T \sim TeV/M_4$, this gives a mechanism to
generate TeV scale effective masses from particles with fundamentally Planckian
masses.

To elaborate on these comments, note that in giving a
four-dimensional description of the physics it is necessary to specify a
reference frame at a definite value of $z$ in terms of which
four-dimensional energies are measured.  The natural frame to use is that
of the Planck brane, as this is  where the 4d graviton bound state is 
supported.  Then if we consider an energy $E_{\rm prop}$ as measured by an
observer at another value of $z$, it will be redshifted so that the energy
in the frame of the Planck brane is $E=\rho E_{\rm prop}/z$.  In
particular, a particle of mass $m$ at rest in the  frame at $z$ will have an
energy $m\rho/z$ relative to the Planck brane, and that will be interpreted
as its four-dimensional mass.

The Lagrangian in this scenario is expected to take the form 
\eqn\tact{\eqalign{S = \int{d^{d+1}}X&\sqrt{-G}({M^{d-1}}\calR-\Lambda 
+ \calL_{\rm stab}) \cr &+ 
\int{d^{d}{x}}\left[
\sqrt{-\gamma(x,\rho_T)}{\cal L}(\gamma(x,\rho_T),\psi)
-\sqrt{-\gamma(x,\rho)}\tau
\right]
\ .}}
Here we denote by $\calL_{\rm stab}$ the Lagrangian of the stabilizing
fields; we assume that beyond stabilizing the brane the don't
qualitatively affect our conclusions.

What is the CFT description of this scenario?  Here we encounter subtleties
beyond the derivation of \deft.  Specifically, the action depends on the
metric at $z=\rho_T$.  In attempting to relate the bulk functional integral
to the boundary CFT we have to confront the non-trivial $z$-dependence of
$\gamma$, and in particular 
give a CFT prescription for computing the metric in the bulk. 
We have not yet found a convincing prescription to derive such off-shell
information from the AdS/CFT correspondence.  

In the absence of such a prescription 
we will consider two approaches to this problem.  The first is to work with
long-wavelength 
excitations of the theory such that the simple scaling approximation
\eqn\gzreln{ \gamma(x,\rho) \simeq {\rho_T^2\over \rho^2} \gamma(x,\rho_T)}
holds.  
%
We use this equation to replace 
$\gamma(x,\rho_T)$ by $\gamma(x,\rho)$ in the Lagrangian for matter on the
TeV brane.  This effectively rescales 
parameters of dimension $\delta$ in that Lagrangian by a factor
$(\rho/\rho_T)^\delta$ ({\it c.f.} \refs{\LyRa}).  Rewriting the
functional integral as in section two produces a 4d effective action
analogous to \deft\ in the Planck brane scenario,
\eqn\tevact{ S_{TeV} = \int d^4x \sqrt{-\gamma} \left[ 
\calL(\gamma,\psi, m\rho/\rho_T) + \calL_{\rm CFT}(\gamma,\chi) + 
\calL_{\rm grav}(\gamma)
-\tau\right]\ .}
Here we have explicitly indicated the rescaling of a typical mass parameter
in the matter Lagrangian.  Again $\calL_{\rm CFT}(\chi)$ 
represents the Lagrangian of
``dark'' CFT matter, and $\calL_{\rm grav}$, given by \sct, is the
gravitational action.  

The simple approximation \gzreln\ fails at short wavelengths,
where the $z$
dependence becomes non-trivial.  This effect can be estimated from the
long-distance expansion of the propagator\GKR,
\eqn\propcorr{\Delta(x,z;x',z')\sim {1\over Rr^{d-2}}\left[{1}+ {R^{d-2}\over
r^{d-2}} + {z^d\over r^d} + {z^{2d}\over r^{2d}}{r^{d-2}\over
R^{d-2}}\right] \left[1+{\cal O}\left({z^2\over r^2},
{R^2\over r^2}\right)\right]\ . }
In $d=4$, the correction due to the last term becomes large at distances
\eqn\corrscal{r\sim \left({\rho_T\over \rho}\right)^{4/3}R\ ,}
or at about 10 Fermi for $\rho/\rho_T \sim TeV/M$.  

In order to understand the origin of corrections at this scale, first
let's recall a similar phenomenon in the large scale compactification
scenario of \ADD.  If one for example considers such a compactification
with two extra dimensions of size $\calo(mm)$, gravitational experiments
performed at shorter scales reveal the six-dimensional nature of
spacetime: the part of the four-dimensional effective Lagrangian
describing the gravitational sector breaks down.  One way of
understanding this is to note that sources with shorter wavelengths than
a millimeter will generically have coupling to the Kaluza-Klein modes
that is comparable to the coupling to the gravitational zero mode;
summing over these modes produces the six-dimensional gravitational
field.  While the gravitational part of the 4d effective action breaks
down, nonetheless the gauge part of the effective action remains
four-dimensional up to scales of order a TeV where gravity itself
becomes strongly coupled.

A similar phenomena occurs here.  At scales given by \corrscal, the
couplings of the TeV brane matter 
to the continuum analogs of the Kaluza-Klein states become
comparable to the coupling to the four-dimensional graviton.  This means
that in the gravitational sector the 4d effective theory fails, but of
course the gauge part of the theory remains four-dimensional up to the
TeV scale.  The stress tensor of the TeV brane matter acts as the source
for these couplings to the Kaluza-Klein modes.

Corresponding statements can be made in the CFT, and will tell us the
form of the corrections to the action \tevact\ that are responsible for
its failure as a 4d effective description.  In particular, we expect
that corresponding to the couplings to the KK modes, a term is induced
in \tevact\ in which there is a direct coupling of the stress tensor of
the TeV brane matter to the stress tensor of the CFT, and by scaling
the coefficient of this should include a factor of $(\rho_T/\rho)^4$.  
Such terms are responsible for the breakdown of the gravitational part
of the 4d effective theory.

A second approach would be to use the holographic renormalization group
\refs{\BaKrhol} to evolve the Lagrangian from the Planck brane to the TeV
brane.  We would expect this to produce similar results, namely a
gravitationally coupled CFT with a cutoff scale $\sim$10 MeV.
We expect the $T_\chi T_\psi$ terms described above to be present in the
Lagrangian at the Planck scale, and then to be rescaled by the
renormalization group flow.  A better understanding along these lines of
the relationship between operators at different $z$ would also clearly
be illuminating for our fundamental understanding of holography in the
AdS/CFT correspondence.

As in the scenario of \ADD, there is a distinction between the scale at 
which the 5d nature of gravity 
becomes important and the scale at which gravity
becomes strongly coupled.  A particularly interesting question is when 
do we expect to be able to manufacture configurations which
would manifest signatures for black holes that we as four-dimensional
observers could see?

Within the context of the TeV-brane scenario, there are again two kinds of
black hole solutions known.  The first is the AdS-Schwarzschild black hole.
The minimum energy to create these should be ${\cal O}(M)$.  A collision of
TeV brane matter with a proper energy of this magnitude is a collision at
the TeV energy scale as measured with respect to the four-dimensional
observer.  However, it appears that such black holes are not bound to the
brane.  In the probe-brane limit, where the gravitational backreaction is
neglected, this is manifest, but even taking into account the small energy
density of the probe brane it seems likely that the binding of the black
hole to the brane will be overcome by the gravitational pull of the black
hole towards the AdS horizon.\foot{There may be interesting transitory
effects -- such as stretching and then recoil of the probe brane -- that we
leave for future investigation.}  While a complete analysis of this
requires detailed investigation of stabilized probe brane scenarios, it
appears that such a black hole will therefore generically fall
towards the AdS
horizon, and that the 4d observer will therefore 
not perceive it as a black hole.  In
the CFT picture, such black holes will be perceived as complex excitations of the
CFT which spread out over time, and it is very unlikely that their
signature can be experimentally 
disentangled from other excitations of the CFT.

The second type of solution is the black hole on the Planck brane.
These are truly perceived as 4d black holes.  However, given the
relation \mreln\ between scales, the minimum energy to create such a
black hole is again of order the Planck energy.  The TeV brane scenario
doesn't seem to allow access to what a 4d observer would perceive as
strongly coupled gravitational dynamics at lower scales.

In fact, notice the following novelty.  A small black hole bound to the
Planck brane 
will not intersect the TeV brane, until its horizon reaches the $TeV^{-1}$ 
size.
Therefore matter moving on the TeV brane may bypass such a 
black hole, in a close analogy to the black hole flybys discussed in
sec.~3.2.  (See fig.~2.)  In other words, 4d observers made of TeV brane
matter have difficulty resolving sub-TeV size black holes.  How is this
interpreted in four-dimensional language?

\ifig{\Fig\Two}{A particle moving on a probe brane can bypass a small black
hole localized on the Planck brane.}{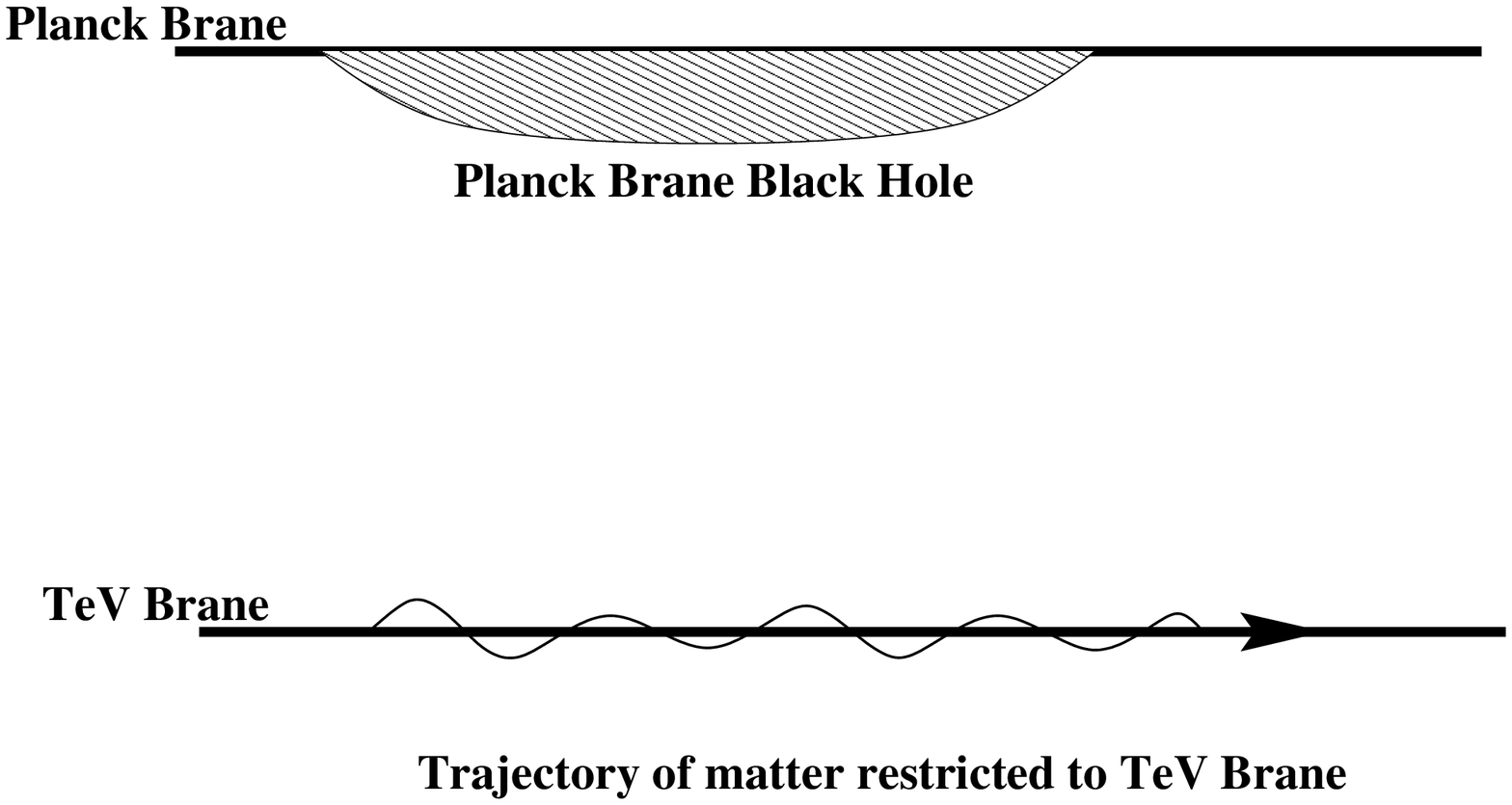}{2.0}

To really study this question requires a detailed model of the
stabilization and the TeV brane matter.  However, a plausible 
answer to this also comes directly from our earlier discussion.  Matter
passing a black hole by moving on the TeV brane should be interpreted in
the 4d perspective as matter smeared out on the TeV scale.  Such matter has
a small probability of probing a black hole with a radius much less than
$1/TeV$.  

\subsec{Terminated AdS scenario}

Another interesting possibility is that AdS is terminated at both ends in $z$.  The
outlines of such a picture was 
suggested in Ref.~\refs{\RSI} with a idealized lower brane
taken to have a finely-tuned negative tension.  

Recent developments in string theory have suggested a concrete means to construct
geometries with similar properties.
Specifically \refs{\KlSt,\MaNu} describe geometries that
terminate at a definite value of $z$.  These geometries do not arise from
negative tension objects, or even singular branes, but rather
are rounded off at the maximal $z$ in a smooth geometry that uses the extra
dimensions of string theory in a non-trivial way.  

Refs.~\refs{\KlSt,\MaNu} do not have a simultaneous microscopic
construction of the analog of a Planck brane.  However, one can envision
building such a model by using Verlinde's geometric realization\refs{\Hver} of the
Planck brane as a piece of a compactification manifold on the ultraviolet
end, and then realize the IR brane as in \refs{\KlSt,\MaNu} or in a variant
of these scenarios producing other low-energy dynamics.  There may of
course be other inequivalent stringy constructions of such
doubly-terminated AdS spaces.  Constructing detailed models of this kind is
an interesting problem for the future.

The models of ~\refs{\KlSt,\MaNu} have explicit gauge theory
duals.  If one constructs a model with a geometric ``Planck brane,'' one
would expect these to be modified at the UV end and depend on the internal
structure of the compactification manifold.  Nonetheless, these gauge
theories should
serve as good effective theories at lower scales, in parallel with our
earlier discussion.

Such a scenario -- or others with a microscopic realization of an IR brane
-- may have very interesting consequences for the observability of strong
gravitational phenomena.  Assume that in such a construction there is a
gauge theory sector that we may think of as being truly localized in the vicinity of
the maximal $z$ which we take to be 
$z\simeq \rho_T$.  This would then
realize what was referred to as matter living on the ``negative tension
brane'' of ref.~\refs{\RSI}.  As explained above, energies at $z\simeq
\rho_T$ are redshifted relative to those at the Planck brane, and so if a
suitable way is found of stabilizing the separation between the branes, TeV
scale scattering corresponds to a proper energy comparable to $M$, the
five-dimensional Planck scale, if the scattering takes place at $z\simeq
\rho_T$.  Thus scattering at this scale should begin to make black holes.
These should be similar to AdS-Schwarzschild solutions, or to the
analogous solutions in the new geometry of ~\refs{\KlSt,\MaNu}.  (For an
explicit formula for the smooth metric in question, see sec.~5.1 of \KlSt.)

In the probe-brane picture, these black holes were expected to fall 
off the brane and into the horizon at $z=\infty$.  Now this is
not possible since the geometry terminates at $z\simeq \rho_T$.  One
expects such  a black hole to undergo approximately geodesic motion in this
vicinity, and ultimately to evaporate.
  
Note that one may achieve a clean separation of scales in situations
where the AdS radius $R$ is larger than the 5d Planck length $M^{-1}$.  In
this situation (which can be achieved by taking large 't Hooft parameter 
$g^2M$ -- here only $M$ is the dimension of $SU(M)$ -- in \KlSt),
there exist black holes larger than the Planck size but smaller than the
AdS radius.  These would have an approximately (five-dimensional)
Schwarzschild description.

In the probe brane picture, the evaporation of 5d black holes was expected
to be nearly exclusively into bulk modes, since the black hole becomes well
separated from the TeV brane and so will not couple to its excitations.
However, in the present picture, the black hole remains in the vicinity of
the analog of the IR brane and this suggests that there is no reason for it
to decouple from the matter modes in this vicinity.  Indeed, in the
idealized ``negative tension brane'' picture, gauge modes on the brane will
directly see the black hole metric.  Therefore, with this assumption, 
on general grounds one expects the black hole to Hawking radiate into all
available modes, including the visible matter sector modes.  As explained
in \refs{\EHMmob}, the radiation in the visible sector is generically expected
to be
important.

This suggests an interesting scenario in which a black hole could be
created at an accelerator operating in the vicinity of the TeV scale.
Assuming the black hole is sufficiently coupled to the visible modes, these
would provide a channel for the Hawking decay and provide an observational
window on this process.  One would observe such an object by looking for
the characteristic approximately thermal spectrum -- with increasing
temperature -- of the Hawking process.

The basic assumptions that could lead to this possibility being realized
are 1) that one has a geometry effectively terminated at a maximal $z$
corresponding to the TeV scale, 2) that one has a description of
visible-sector matter localized to the vicinity of this maximal $z$, and 3)
that black holes near the maximal $z$ couple to the visible sector.
Whether these assumptions will hold in models based on the ideas of
\refs{\KlSt,\MaNu} remains to be seen, but they plausibly do, and 
there may also be other
models with these properties, for which the
creation and visible-sector decay of TeV-scale black holes seems a generic
prediction.

\newsec{Conclusions}

This paper has investigated the interplay between the four- and
five-dimensional descriptions of the physics of warped compactifications.
In the simplified example of the RSII scenario, at distances long as
compared to the AdS radius $R$ there is a four-dimensional effective
description of the dynamics given by observable brane matter coupled
gravitationally to a sector described by a conformal field theory.  At
shorter distances the derivation of this description fails.  One expects
similar 4d effective descriptions for other warped compactification
scenarios.

One element of the correspondence between the 4d and 5d descriptions is
supplied by the computation of the 4d stress tensor corresponding to a 5d
matter distribution.  At the linear level we have given a formula for this
stress tensor.  We have also investigated an amusing scenario that
illustrates the interplay between the 4d and 5d descriptions, that of a
particle passing a black hole through the fifth dimension, with a
corresponding 4d description in terms of a matter distribution expanding
into a shell larger than the black hole.

Finally, we have explored situations in which strong gravitational dynamics
may give important modifications to the 4d description.  In particular, in
scenarios where the hierarchy is addressed by visible matter being
effectively localized to a large $z$ in AdS space, one potentially has
access to strong gravitational dynamics such as black hole formation at TeV
energy scales.  In probe brane scenarios this may not lead to observable
effects since the resulting black hole seems to rapidly decouple from the
visible sector by falling off the brane, but scenarios with AdS terminated
at this maximal $z$ show much more promise as such a black hole should stay
localized in the vicinity of the maximal $z$.  This leads to the
possibility of creation and observable Hawking decay of a black hole in the
vicinity of the TeV scale.  It would be particularly interesting to find
extensions of the work of \Hver\ and \refs{\KlSt,\MaNu} which give explicit
string theory realizations of such terminated AdS scenarios.

\bigskip\bigskip\centerline{{\bf Acknowledgments}}\nobreak

The authors wish to thank N. Arkani-Hamed, 
G. Horowitz, L. Randall, H. Verlinde, and E. Witten for
valuable conversations.  Parts of this work were carried out at Caltech,
and the Caltech/USC Center for Theoretical Physics,
whose support and hospitality are gratefully acknowledged, at the Aspen
Center for Physics, and at the Univ. of Colorado, Boulder. The work of SBG was
partially supported by DOE contract DE-FG-03-91ER40618, and that of EK by
DOE contract DF-FC02-94ER40818.

\appendix{A}{The retarded Green function}

In this appendix we describe some properties of the scalar Green function
for the RSII geometry.  This was given in \GKR\ and takes the form
\eqn\finalG{\eqalign{
\Delta_{d+1}(x,z;x',z')& ={i\pi\over 2R^{d-1}}(zz')^{d \over 2} \int  {d^d p 
\over
(2\pi)^d}e^{ip(x-x')} \cr
& \times \left[{J_{{d \over 2}-1}(qR) \over H_{{d \over 2}-1}^{(1)}(qR)}
H_{d \over 2}^{(1)}(qz) H_{d \over 2}^{(1)}(qz') 
-J_{d \over 2}(qz_<)H_{d \over 2}^{(1)}(qz_>)\right]\ .}}
For the following discussion it is most convenient to use the $z$
coordinate, related to $y$ by \yzreln.

The scalar propagator with $d=4$ and 
one point on the boundary is given by\GKR
\eqn\onbrane{ \Delta_{4+1}(x,z;x',R) = \left({z\over R}\right)^{2} \int
{d^4p\over  (2\pi)^4} e^{ip(x-x')} {1\over q} {H_{2}^{(1)}(qz) \over 
H_{1}^{(1)}(qR)}\ ,} 
where $q^2=-p^2$.  As in eq. \gammadef, let us define a function $F$,
\eqn\gammadefz{\Delta_{4+1}(R,x;z',x') = {z'^3\over R} \partial_{z'} 
\left[{
F_{z'}(R;x-x') \over z'^2}\right] \ .} 
Hence, $F$ is given as Fourier transform of Hankel functions,
\eqn\Fhankel{F_{z'}(R;x-x') = 
-{z'\over R} \int
{d^4p\over  (2\pi)^4} e^{ip(x-x')} {1 \over q^2} {H_{1}^{(1)}(qz') \over 
H_{1}^{(1)}(qR)}\ .}

In our conventions (given by \branegf) the Feynman propagator is
\eqn\Feynprop{\Delta_F(X,X') = -i \left[\theta(t-t') \Delta^{+}(X,X') +
\theta(t'-t) \Delta^{-}(X,X') \right]\ ,}
where $\Delta^+$ is the Wightman function $\langle \phi(X)\phi(X') \rangle
$ and $\Delta^-$ is its hermitian conjugate.
The retarded Green function is defined as
\eqn\DefRet{\Delta_R(X,X') = -i \theta(t-t') \left[ \Delta^{+}(X,X') -
\Delta^{-}(X,X') \right]\ ;}
this manifestly vanishes for $t<t'$ and can easily be shown to obey \branegf.
We therefore find that 
\eqn\ftoret{\Delta_R(X,X') = 2 {\rm Re} \Delta_F(X,X') \theta(t-t')\ .}

In order to compute the asymptotic retarded Green function, note that 
in the long distance approximation ($qR << 1$), $F$ reduces to the
following form,
\eqn\Flong{F_{z'}(R;x-x') \approx -{{\pi iz'} \over 2} \int
{d^4p\over  (2\pi)^4} e^{ip(x-x')} {1 \over q} H_{1}^{(1)}(qz')\ .}
We then perform a
Euclidean rotation on the above integral, giving
\eqn\FlongEuc{\eqalign{F_{z'}(R;x-x') & \approx {{\pi i}z' \over 2} \int
{d^4p\over  (2\pi)^4} e^{ip(x-x')} {1 \over p} H_{1}^{(1)}(ipz') \cr
& \approx {i \over {4 \pi^2}} {1 \over {(x-x')^2 + z^2}}\ .}}
The Feynman prescription is then to replace $(x-x')^2$ with $(x-x')^2 +
i\epsilon$.  Making this replacement, and taking the real part gives
\eqn\Fretard{F_{z'}^{Ret}(R;x-x') \approx {1 \over 2\pi} \delta((x-x')^2
+z'^2)\theta(t-t')\ ,}        
which finally yields the retarded scalar propagator,
\eqn\Retprop{\Delta_{4+1}(R,x;z',x') \approx {1 \over \pi R} \left[z'^2
\delta^{\prime}((x-x')^2 +z'^2) - \delta((x-x')^2 +z'^2) 
\right])\theta(t-t')\ .}

\appendix{B}{The infrared limit}

As we saw in the text, our calculation of the effective source on the
boundary produces an expanding shell in the infrared limit.  This is true
for scalar, vector, and graviton fields.  This is also the result that
Horowitz and Itzhaki found in \HoIt, using the boundary conditions
appropriate for infinite AdS rather than the brane boundary conditions
\bcond.  In this appendix we sketch the relation between the calculations.
For simplicity we only treat the scalar case although the derivation
extends to the other cases.

In our calculation with brane boundary conditions, we solve the bulk
equation
\eqn\bBeqn{ \sq_{d+1} \phi = T\ ,}
with the Neumann condition
\eqn\phibC{\partial_{n}\phi\vert_{z=\rho}=0\ .}
Here $T$ is the scalar source, in
the text given by the falling particle, and $\partial_{n}$ denotes the
normal derivative.  The effective boundary source is
found by restricting this solution to the boundary and computing its
Laplacian:
\eqn\effsour{J = \sq_d \phi\vert_\partial\ .}

Another way to get the same solution is to 
solve \bBeqn\
subject to the Dirichlet boundary condition 
\eqn\adsbc{\phi\vert_\partial = \varphi\ .}
The solution is given in terms of the Dirichlet Green function as 
\eqn\dirsoln{\phi(X) = \int dV' \Delta^D_{d+1}(X,x') T(X') + \oint_\partial dn'
\partial_{n'} \Delta^D_{d+1}(X,X') \varphi(x')\ .}
The effective boundary action for $\varphi$ is computed by evaluating the
$d+1$-dimensional action of this solution, which using the bulk equation of
motion becomes
\eqn\phiact{S[\varphi] = -{1\over 2} \oint_\partial dn' \varphi \partial_{n'}\phi\
.}
The boundary equation of motion for $\varphi$ then states
\eqn\phieom{\partial_n\phi\vert_\partial =0\ .}
Thus a solution of the Dirichlet boundary problem such that the boundary
field satisfies the boundary equations of motion corresponds to a solution
of the Neumann boundary problem.

In the latter approach the effective boundary source can be read off from
the boundary equation of motion.  Inserting the second term of \dirsoln\
into
\phieom\ gives the kinetic operator acting on $\varphi$, which becomes
$\sq_d$ in the long distance limit.  Thus in this limit \phieom\ states
\eqn\effsource{J=\sq_d\varphi \propto \partial_z \phi_D|_\partial\ ,}
where $\phi_D$, the first term in \dirsoln, is the solution to \bBeqn\ 
with Dirichlet
boundary conditions, $\phi_D\vert_\partial =0$.  In the limit as the cutoff
is removed, $\rho\rightarrow0$, this corresponds to the desired solution in
infinite AdS.  And aside from a rescaling, $\partial_z \phi_D$ corresponds to the
source on the boundary, which if we had been discussing the metric would be
the boundary stress tensor of \HoIt.

It is also possible to check the relationship to \HoIt\ directly, by acting
with $\sq_d$ on \scalsoln, using the eq. \branegf\ to eliminate the
d-dimensional laplacian in favor of $y$ derivatives, and then using the
fact that in the infrared limit \finalG\ is the standard bulk propagator
plus a $y$-independent piece which therefore doesn't contribute.

\listrefs
\end